# Insight into Non Linearly Shaped Superconducting Whiskers via Synchrotron Nanoprobe

**Stefano Cagliero**[1], **Elisa Borfecchia**[1], **Lorenzo Mino**[1], **Leandro Calore**[1], **Federica Bertolotti**[1], **Gema Martinez-Criado**[2], **Lorenza Operti**[1], **Angelo Agostino**[1], **Marco Truccato**[3], **Petre Badica**[4] **and Carlo Lamberti**[1]

[1] Department of Chemistry, NIS Centre of Excellence, INSTM Unit, University of Turin, Via P. Giuria 7, 10125, Turin (Italy)
[2] European Synchrotron Radiation Facility, 6 rue Jules Horowitz, BP220, F-38043, Grenoble CEDEX (France)
[3] Department of Physics, NIS Centre of Excellence, CNISM UdR University of Turin, Via P. Giuria 1, 10125, Turin (Italy)
[4] National Institute of Materials Physics, Atomistilor 105bis, 077125, Magurele (Romania)

E-mail: marco.truccato@unito.it

**Abstract.** We managed to synthesize non-linear YBa2Cu3Ox whiskers, i.e. half loops or kinked shapes, which are promising candidates for solid-state devices based on the intrinsic Josephson effect and with improved electrical connections. We report on a complete characterization of their structural properties via synchrotron nanoprobe as well as laboratory single-crystal diffraction techniques. This investigation allowed us to fully disclose the growth mechanism, which leads to the formation of curved whiskers. The superconducting properties are evaluated in comparison with the straight counterpart, revealing a strong functional analogy and confirming their potential applicability in superconducting electronic devices.

## 1. Introduction

Since the discovery of high-$T_c$ cuprate superconductors, much attention has been dedicated to the growth of single crystals for fundamental research and electronic applications. In this framework, since a few years ago, $Bi_2Sr_2CaCu_2O_{7+x}$ (Bi-2212) and $YBa_2Cu_3O_x$ (Y-123) whisker-like crystals attracted significant attention because of their very good crystallinity, peculiar dimensions with micrometric cross sections, leading to high aspect ratios and outstanding superconducting properties.
The successful growth of Y-123 whiskers has been reported in more recent years than the well-known Bi-2212 family, but a good level of characterization of their superconducting properties has already been achieved [1, 2].
A series of recent publications shows that high-$T_c$ whiskers, exploiting the intrinsic Josephson effect and related phenomena, can be successfully employed in the fabrication of micro/nano-devices, such as: THz emitters/sensors, micro-SQUIDs' and quantum bit-computing based on Macroscopic Quantum Tunneling phenomena [3-7]. From the applications point of view, the possibility of obtaining superconducting crystals with curved shapes, maintaining the same superconducting properties of the straight ones, would open the way to the realization of novel and interesting solid-state devices. For instance, solid state micro-coils with high critical currents could be applied in the generation of extreme magnetic forces, while the exploitation of the intrinsic Josephson effect could take advantage from the introduction of diverse crystal geometries, like half loops or kinked shapes. Moreover, the availability of curved whiskers would allow the fabrication of systems with improved electrical connections [8, 9].

While the production of curved outlines has been reported in the past for metal whiskers [10] and simple oxides nanowires and nanobelts [11, 12], providing a characterization of the structure and of the properties related to their shape, literature shows very few examples of non-linearly shaped Y-123 or Bi-2212 crystals, like for instance the ones obtained as a secondary product of Y-123 sputtering process [13]. In the present paper we report on the growth of Y-123 whiskers bent along the growth direction and we discuss a structural comparison with the straight ones obtained by nano-probe and single crystal X-ray diffraction techniques along with their superconducting behaviours.

## 2. Experimental methods

### 2.1 YBCO whiskers synthesis

YBCO whiskers were grown by means of the following method:[3] powders with nominal cationic composition of $Y_1Ba_2Cu_3Ca_1Te_{0.5}$ were prepared by solid-state reaction from individual component oxides by using highly pure precursors of $Y_2O_3$ (99.99%), $BaCO_3$ (99.999%), CuO (99.9999%), $TeO_2$ (99.995%) and $CaCO_3$ (99.9999%) (Sigma-Aldrich, Germany). The precursor powders were thoroughly mixed and calcined at 900°C in alumina crucibles for 10 hours in air with three intermediate grindings. The powders were then pressed into pellets of about 13mm in diameter and 2mm in thickness. The pellets, distributed on $Al_2O_3$ boats, were heated up to 1005 °C ($T_{max}$) with a constant rate of 5°C/min, dwelling at $T_{max}$ for 5 hours with a controlled oxygen flow of 0.15 l/min. The cooling ramp was carried out at a rate of 1°C/h, down to the ending temperature $T_{end}$=925°C; finally the pellets were furnace cooled to room temperature.

This synthesis leads to both straight and curve shaped YBCO whiskers, whose morphology was observed by scanning electron microscopy (SEM).

### 2.2 Laboratory XRD measurements

Single crystal X-ray diffraction (XRD) data were collected using a Gemini R Ultra diffractometer operating at 50 kV and 40 mA, with a graphite-monochromatized Mo Kα (λ= 0.71073 Å) source radiation and using an ω-scan technique (Δω = 1.0°). CrysAlis PRO software has been used [14] for data collection and reduction (peaks intensities integration, background evaluation, cell parameters and space group determination).

### 2.3 Synchrotron nanoprobe setup

In order to gain a deeper insight into the crystalline nature of the curved samples we made use of an intense hard X-ray nanobeam available at the ID22 beamline of the European Synchrotron Radiation Facility (Grenoble).

Due to the recent development of the hard X-ray nanostation (see Fig. 1a), only few studies have been reported in different research areas (e.g. biomedical [15-18], earth, and environmental sciences [19, 20], as well as nanomaterial research [21, 22]).

Located at 64 m from the undulator source, the nano-focusing optics consists of two graded multilayer coated surfaces mounted in crossed Kirkpatrick-Baez configuration [23]. A similar mirror pair has been already employed by our group in previous microbeam experiments [24-27]. Here, the X-ray lens is composed of a 112 mm-long mirror focusing at a distance of 180 mm from the centre of the mirror in the vertical direction and a 76 mm-long mirror with a 83 mm focusing distance in the horizontal direction. Four actuators (μ-Focus picomotors) bend the flat polished mirrors (CoastLine Optics) into the elliptical figures required for imaging the X-ray source. Both arms of each bender are equipped with linear encoders (Mercury 3500). This design provides reflectivity of 73% at 17 keV and 74% at 8 keV. A virtual source is typically created in the horizontal direction based on the high heat-load slits (depending on the spatial resolution and photon flux required by the experiment, from 10 up to 25 μm). The first mirror plays both the role of vertical focusing device and monochromator, resulting in a very high flux of about $5 \times 10^{12}$ photons/s and medium monochromaticity of $\Delta E/E = 10^{-2}$. The larger incidence angle of the multilayer increases significantly the acceptance. Thus, it can be exploited in both pink and monochromatic beam operations based on Bragg diffraction and total external reflection modes, respectively. The pink beam approach uses the multilayer configuration to increase beam divergence (numerical aperture), whereas the second strategy, optimized for X-ray absorption spectroscopic acquisitions, relies on grazing incidence to provide a monochromatic beam flux of about $5 \times 10^{10}$ photons/s. Invar as material choice for the benders has improved the thermal stability, and in particular, the stability of the curvature of the elliptically shaped mirrors. In order to avoid amplification of the mechanical vibrations, all instrumentation is mounted on a single table consisting of a concrete base and a 4.4 m long granite bed. In addition, the standard in-hutch air conditioner is also replaced by a complete air-exchange system, with associated chicanes and a vestibule entrance to reduce hutch contamination, allowing a thermal stability of $\pm 0.1\,°C$.

For our nano-XRD measurements, apart from the X-ray focusing optics located in the nanoimaging station, a horizontally deflecting mirror to remove the high energy X-rays from the incident beam and a double Si (111) crystal monochromator, both placed in the optics hutch, were part of the experimental setup. XRD experiments were carried out using a sample-detector distance of 113 mm, a 17.1 keV X-ray nanobeam with dimensions (evaluated by the knife-edge scan method [28] at the focal plane) of 107 nm vertically and 152 nm horizontally, working in pink beam mode ($\Delta E/E \approx 1\%$).

YBCO whiskers were fixed on apposite plastic holders with adhesive tape, making use of an optical microscope to select and locate the crystals to be measured. Each whisker was aligned with the beam direction normal to its largest surface, typically coinciding with the *ab*-plane [2]. The sample holder was mounted on a computer-controlled piezo-positioner stage coupled with a microscope, which achieved sample positioning with 0.1 µm resolution. The diffraction patterns were collected using a FreLon CCD camera with 14-bit readout, 2048 x 2048 pixels; pixel size of 14x14 µm$^2$. XRD data were examined using the Fit2D software, after correcting for CCD intrinsic spatial distortions and calibrating with LaB$_6$ standard. The average diffraction image on all sampled points used for reflections indexing was obtained using the Fit2D software, while the indexing procedure was performed by comparing the average XRD image with the theoretical diffraction pattern simulated from YBCO orthorhombic structure.

*2.4 Magnetization measurements*

Magnetization measurements were performed by means of a SQUID magnetometer (Quantum Design ®, MPMS-7T). The data were collected at a fixed field of H=100 Oe and with the Zero Field cooling technique. To obtain significant signal to noise ratio, we used eleven aligned straight whiskers. Conversely, a single curved whisker was sufficient, being the volume of curved crystals much larger with respect to the straight ones. The crystals were fixed with conventional plastic straw and vacuum grease (Apiezon M ®).

**3. Results and discussion**

Examples of scanning electron microscope (SEM) images of the most commonly found shapes are illustrated in Fig. 1b, and c, where straight and curved samples are shown, respectively. The percentage of curved whiskers obtained in each synthesis exhibits small fluctuations around a typical value of 35%, owing to local variations in the microscopic growth environment.

Table 1 summarizes the cell parameter values determined on three straight and three curved YBCO whiskers collected from the same synthesis batch. The reported values agree with previously published cell parameters obtained for bulk YBCO samples, which are normally in the range 3.815-3.835 Å for the *a*-axis, 3.865-3.888 Å for the *b*-axis, and 11.68-11.74 Å for the *c*-axis [14, 29-31]. To the best of our knowledge, no evaluation of the *a*, *b* and *c* lattice parameters has ever been performed for YBCO whiskers, except for the above mentioned crystals obtained as a secondary product of sputtering technique [13]. From Table 1 evidently a certain degree of fluctuation is present among all cell parameters. However, no significant trend is recognized, which could discriminate between straight and curved specimens. Likely the observed fluctuations arise from the inhomogeneous conditions in the precursor pellets, resulting in slightly different amounts of Ca, Te and O incorporated during the synthesis process.

**Table 1.** Cell parameters of three straight and three curved whiskers, as obtained by single-crystal XRD measurements. The reference compound crystallizes in an orthorhombic *Pmmm* space group.

|  | Straight Whiskers | | | Curved Whiskers | | |
| --- | --- | --- | --- | --- | --- | --- |
|  | $S_1$ | $S_2$ | $S_3$ | $C_1$ | $C_2$ | $C_3$ |
| a (Å) | 3.8281(4) | 3.8149(3) | 3.8155(2) | 3.8237(4) | 3.8299(4) | 3.8230(2) |
| b (Å) | 3.8703(4) | 3.8781(3) | 3.8744(4) | 3.8701(4) | 3.8658(4) | 3.8614(6) |
| c (Å) | 11.6837(9) | 11.6789(6) | 11.6772(6) | 11.682(1) | 11.6745(8) | 11.6705(2) |
| Reflections Collected | 2772 | 2605 | 2590 | 1322 | 3466 | 2267 |
| Unique Reflections | 310 | 298 | 298 | 293 | 309 | 297 |

The measurements suggest a single crystal nature for both straight and curved whiskers. This observation leads us to assume a structural picture for the curved crystals analogous to that of the straight ones: the cells are always aligned along the initial growth direction, and only their stacking laterally translates during the growth (Scheme (2) in Fig. 2c). A possible alternative explanation would involve multiple domains twinned in the *ab*-plane (Scheme (1) in Fig. 2c) with the crystalline cell rotating along the whisker length. Unfortunately, standard single-crystal diffraction provides global information averaged on the whole sample volume, being unable to detect the coexistence of one main domain with multiple much smaller secondary domains that are not large enough to produce a significant diffraction signal.

In order to achieve deeper insights into this issue, a spatially resolved approach with high sensitivity (signal-to-noise ratio) is therefore crucial. High Resolution Transmission Electron Microscopy (HRTEM) is hardly applicable in this specific case because a systematic sampling along the whole whisker curvature (30-100 µm) would be very difficult to be realized; moreover, the thickness of the sample would require a non-trivial and possibly invasive thinning process. Alternatively, to validate the hypotheses resulting from the single-crystal XRD measurements, these limitations can be overcome by using an intense hard X-ray nanobeam, like the nanoprobe available at the ID22 beamline of the European Synchrotron Radiation Facility (Grenoble) [23].

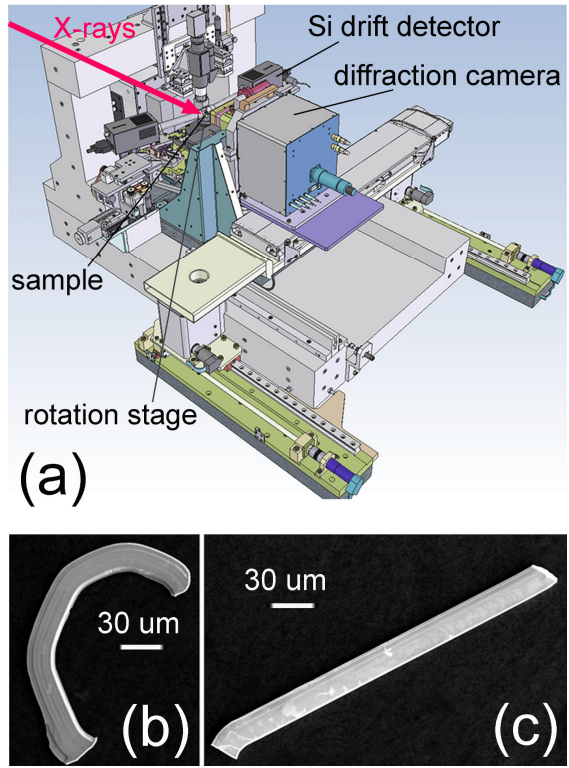

**Fig. 1** (a) Overview of the experimental arrangement of the nanoimaging station of ID22, showing the incoming X-ray beam direction, the sample position and the available detectors for X-ray fluorescence and diffraction measurements. (b), (c) SEM micrographs of representative curved and straight $Y(Ba,Ca)_2Cu_3O_{7-x}$ whiskers, respectively.

Fig. 2a summarizes the results obtained by synchrotron nano-XRD investigation on a single curved YBCO whisker, representative of the general behavior observed for all the samples. The positions indicated and labeled using numbers from 1 to 5 in the SEM image correspond to the points along the whisker curve where the diffraction patterns were recorded. An additional point was sampled near the center of curvature of the bowed whisker, in a thinner region of the crystal (point 6 in Fig. 2a). The six reported frames are only a selection among the 50 diffraction patterns collected along the whole sample curvature. The average image of the six frames is reported in Fig. 2b, with assigned *hkl* reflections highlighted by violet circles. The pattern clearly describes the crystal $a*b*$-plane in the reciprocal space.

Comparing the information from the average image with each single frame, it is possible to notice that all the frames are providing the same diffraction pattern while moving along the crystal profile: such outcome is a clear evidence of the single crystal nature of the curved YBCO whisker.

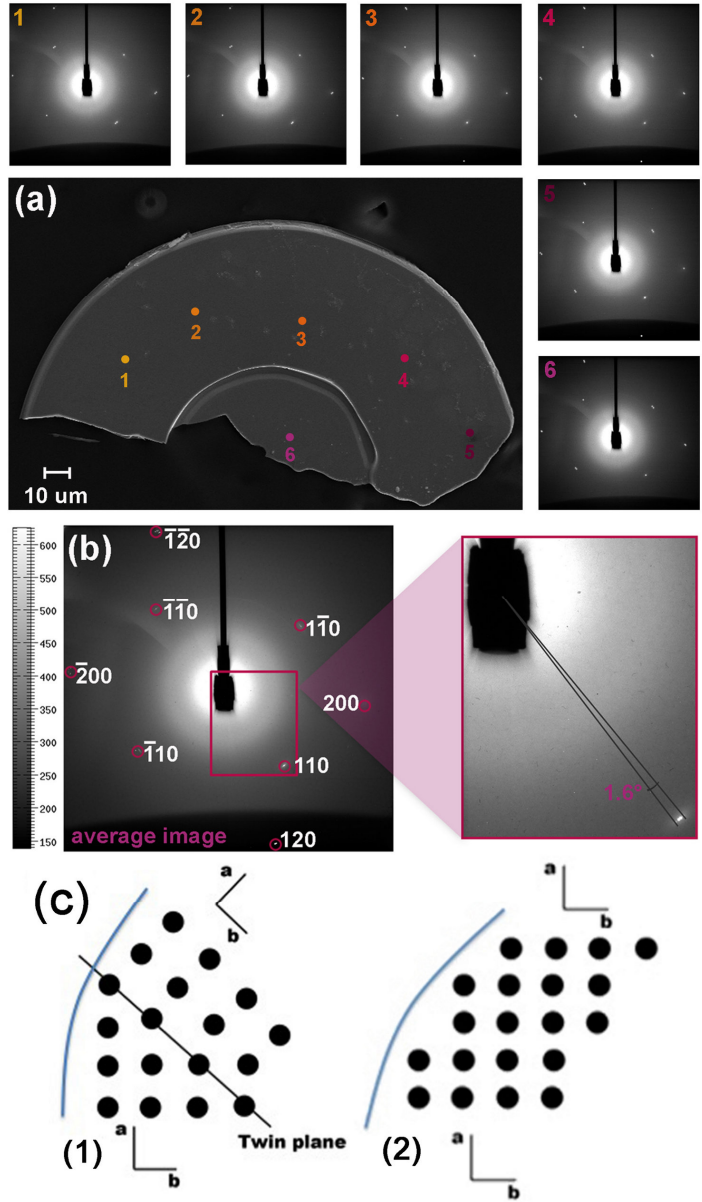

**Fig. 2** Results obtained by synchrotron nano-XRD investigation of a curved YBCO whisker. (a) SEM image of the analyzed semicircular shaped whisker. The spatial positions where diffraction patterns were collected are indicated by numbers from 1 to 6. The corresponding XRD 2D-patterns (after correction for CCD distortions) are reported around the SEM image. For clarity, the intensity of diffraction spots has been graphically enhanced. (b) Indexed reflections on the averaged image of XRD frames 1-6 in (a); on the right, a magnification of the (1 1 0) reflection is reported. (c) Schematics of possible cell arrangements resulting in a non-linear whisker shape: (1) bowed profile originated from a continuous succession of growth twins in the *ab*-plane, and (2) curvature derived from a lateral translation of crystal cells, always aligned along the initial growth direction.

The determination of *a* and *b* cell parameters from the nano-XRD data resulted in axes values compatible with those reported in Table 1, but affected by higher experimental errors owing to the set-up not optimized for cell parameter determination (only one frame per point was collected). Within the experimental error no appreciable fluctuation was observed along the 50 sampled points. By closely looking at each single frame and at the average image (Fig. 2b, magnified view of the (1 1 0) reflection), it is possible to observe that some reflections are not point-like, but consist in double spots.

This feature is due to the (1 1 0) – (1 -1 0) twinning that usually takes place in Y-123 single crystals during the tetragonal-orthorhombic transformation at high T [31, 32]. The presence of multiple twin domains has been studied on YBCO film samples by synchrotron radiation [14], resulting in double spots where one twin set dominates (1 1 0) or (1 -1 0) and quadruple spots at the twin domain walls. In our experiment at ID22 beamline only double spots were observed because of both the much smaller beam size. On the other hand, by taking into account frames from the single crystal diffraction measurements, which is an average measurement technique, quadruple twin features can be noticed on high angle reflections. As an example, we have reported in Fig. 3a, the (2 0 16) reflection, which is comparable with the results obtained by Caldwell *et al.* [14] along with the magnification of (1 1 0) double spot in Fig. 3b, also shown in Fig. 2a (frame 1).

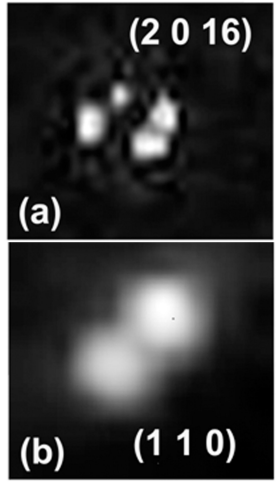

**Figure 3.** Magnified images of twinning features of YBCO whiskers. Panel (a), (2 0 16) reflection from single crystal laboratory diffraction measurement: four spots are present due to volume averaged character of the technique. Panel (b), (1 1 0) reflection from nano-beam X-ray diffraction: only two spots are observed due to the much smaller beam size.

The angle enclosed by the double spot shown in Fig. 3b is about 1.6°, which is in qualitative agreement with what expected for (1 1 0) twinning from the determined values of the lattice parameters. Such a small angle is clearly not ascribable to a rotation of the crystal cell, as shown in the model 1 of Fig. 2c. Indeed, this rotation angle, in the case of a continuous succession of growth twins in the *ab*-plane as the origin the crystal bending (see Fig. 2c), is expected to be about 120° between points 1 and 5, being linked to the crystal habit. We observed similar features on all whiskers, independently on their shape (straight or curved), which further evidence that the twin features are not correlated to the whisker curvature. Therefore, these results further prove that the curved shape of such crystals should result from the recently proposed mechanism based on a shape and size changing, moving and multiple growth interface (SCMMGI) [33].

Finally, we have compared the superconducting properties of curved and straight whiskers by means of magnetization measurements: Fig. 4(a) displays the magnetic moment versus temperature raw data both for a single curved whisker and for a collection of 11 widely spaced, straight whiskers. Apart from the different scales of the magnetic signal related to the different sample volumes and shapes, nearly identical critical temperatures were obtained of $T_c$=79 K, similar to the ones previously reported for YBCO straight whiskers [1, 2].

We have also performed a magnetization analysis to take into account the different sizes of curved and straight crystals. SEM observation revealed that the curved whisker was about 350 μm long, 60 μm wide and 2 μm thick, while the average sizes of the straight whiskers were 200 μm in length, 10 μm in width and 1 μm in thickness, corresponding to a total volume for the 11 straight crystals that was about half of that of the curved whisker. Since the magnetic field was applied along the thickness direction and the length was much greater than the other sizes in all of the measured crystals, the approximation of an infinite rectangular bar in transverse field was applied for the calculation of the magnetometric demagnetizing factor, resulting in $N_m$= 0.96526 and $N_m$ = 0.89958 for the curved and for the average straight whiskers, respectively [34]. The corresponding results for the volume magnetic susceptibility $\chi$ are shown in Fig. 4(b): in spite of the rough approximation consisting in considering all the straight whiskers as equal to the average straight one, it is clear that both samples share the same fully diamagnetic behaviour. Minor differences are visible only near the transition region, which are probably due to the dispersion of the size distribution for the straight whiskers.

These results are in perfect agreement with the structural analyses, proving the striking similarity of the two types of crystals also from the superconducting point of view.

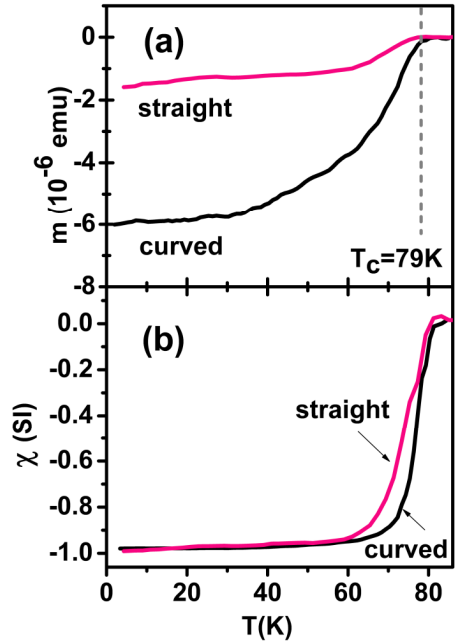

**Fig. 4.** Panel (a): Magnetic moment (*m*) versus temperature (*T*) at *H*=100 Oe for straight and curved whiskers. The critical temperature ($T_c$) is indicated by the dashed gray vertical line. The background signal of the bare sample holder was subtracted in both cases. Panel (b): Volume magnetic susceptibility $\chi$ in SI units for both the curved and the straight whiskers as obtained from their sizes and the corresponding demagnetizing factors (see text).

## 4. Conclusions

In summary, we have demonstrated the close analogy between straight and curved YBCO crystals from both the structural and the superconducting points of view. On the basis of nano-diffraction investigation, we confirm a structural arrangement with the cell oriented along the starting growth direction, laterally shifting in its stacking while the growth proceeds, thus giving rise to a macroscopic curvature. Possible roles of twinned domains in the bowed shape formation have been ruled out. Moreover, a complete cell parameter characterization has been provided, with new evidences of structural fluctuations related with local inhomogeneities at the surface of the precursor pellet. Concerning the superconducting properties, a $T_c$ value of 79 K and full Meissner state have been found for both systems by means of magnetization measurements. These results set a starting point for future control and exploitation of non-linear shaped crystals in superconducting electronic devices.

**Acknowledgments**

PB acknowledges financial support from UEFISCDI (Romania), project PNII PCCA 138/2012. SC, EB, LM, LC, AA, MT and CL acknowledge financial support from project ORTO11RRT5 in the framework of Progetti di Ricerca di Ateneo - Compagnia di San Paolo-2011.